\documentstyle[epsf,11pt]{article}

\newcommand{\vr}[1] {{\vec r_{#1}}}
\newcommand{\vR}[1] {{\vec R_{#1}}}
\newcommand{\cmod}[1] {|{#1}|^2}

\begin{document}
\title{Bio-Polymer Hairpin Loops Sustained by Polarons}

\author{
B. Chakrabarti\thanks{e-mail address:  buddhapriya.chakrabarti@durham.ac.uk},\,
B.M.A.G. Piette\thanks{e-mail address: B.M.A.G.Piette@durham.ac.uk},\,
W.J. Zakrzewski\thanks{e-mail address: W.J.Zakrzewski@durham.ac.uk}
\\
Department of Mathematical Sciences, University of Durham, \\
Durham DH1 3LE, UK\\
}
\date{}
\maketitle

\begin{abstract}
We show that polarons can sustain loop-like configurations in flexible
bio-polymers and that the size of the loops depend on both the 
flexural rigidity of the polymer and the electron-phonon coupling constant.
In particular we show that for single stranded DNA (ssDNA) such
loops can have as little as 10 base pairs. For polyacetylene the shortest 
loop must have at least 12 nodes. We also show that these configurations
are very stable under thermal fluctuations and can facilitate the formation of
hairpin-loops of ssDNA.  
\end{abstract}

\section{Introduction}
Conformational transitions of biopolymers as a
result of the coupling between the electronic and elastic degrees of
freedom are important for understanding
native states of globular proteins and secondary structures of
biopolymers such as DNA and RNA. In an attempt to understand toroidal
states of DNA  the globule-coil transition for semi-flexible polymers
in poor solvents has been explored  using Brownian dynamics
simulations~\cite{Schnurr00,Schnurr02}. The intermediate states
arising in these systems have also been
classified~\cite{Schnurr00,Schnurr02}. However the  collapse transition in
polymers induced by polarons has been less explored~\cite{Mingaleev02}.

Polarons are the result on the interaction between a free electron in the
conducting band of a polymer chain and the phonons of that chain.
They were discovered by 
Davydov~\cite{Davydov1973,Davydov1979,Davydov1981} 
who proposed them as a
mechanism to explain how energy can be transported along
alpha-helices in living cells.  

In this paper we explore the possibility of polaron induced polymer
loop formation and  stabilisation arising in semi-flexible chains. Our
model is similar to the model proposed  by Mingaleev et
al.\ \cite{Mingaleev02} who generalised the original model of
Davydov~\cite{Davydov1973,Davydov1979} by incorporating long ranged
electron-phonon  interactions. In their work Mingaleev et
al. showed that at zero temperature  polarons can
induce a spontaneous bend in a straight chain if the bending modulus
is less than  a critical threshold. A careful examination of the model
however reveals that realistic polymers \textit{e.g.}  DNA and
polyacetylene are more rigid having their bending modulus twice and
twenty times above the  threshold value respectively. 
Therefore though interesting
from a theoretical point of view, the spontaneous bending on polymers 
induced by polarons is limited in  scope when applied to physical systems.  

In a recent paper~\cite{CPZ12} we showed
that the Mingaleev et al.\ model  can explain spontaneous polaron
transport on a chain having a bending gradient, \textit{e.g.}
alpha-helices  of light harvesting proteins. In this case, the bending
of the chain is generated by the natural folding  of the protein which
can induce a spontaneous polaron displacement. We showed that with the
polymer configuration frozen in, the polaron spontaneously
accelerates along the bending gradient, and gets reflected across
sharply kinked junctions. Further we showed that at finite
temperatures the polaron undergoes a biased random walk to a region
of high curvature. 

While polarons are not able to induce spontaneous conformational transitions 
in DNA and polyacetylene, on account of their rigidity, 
they might sustain a folded configuration that might have
been formed by other means \textit{e.g.} thermal fluctuations, or
mechanical stress. This is particularly true for ssDNA  whose bending
modulus is only twice as large as the threshold value for spontaneous bending.
This is what we are investigating in detail in this paper which 
is organised as follows.

In  section \ref{Sec:Model} we review the Mingaleev et al.~\cite{Mingaleev02} 
model . In section \ref{Sec:Loop} we study loop configurations in  
which the last two
nodes of the chains are held together by a polaron. We extend this
analysis to  study hairpin-loop configurations in
\ref{Sec:Hairpin} for which the two opposite ends of the chain  run
parallel to each other, while the loop links the parallel strands
together. Finally we show that one can estimate analytically the value
of the parameters for which loops can be formed in
section\ref{Sec:Results-Analytical}. In the last two sections
\ref{Sec:ssDNA}  and \ref{Sec:Poly} we look in some detail at
loop and hairpin-loop configurations for both  single stranded DNA and
polyacetylene and we show that the polarons, in these two systems, are
very  stable and that they can facilitate the formation of
hairpin-loops. 

\section{Model}
\label{Sec:Model}
The model proposed by Mingaleev~\cite{Mingaleev02} is described by the 
Hamiltonian 
\begin{eqnarray}
H = \sum_{n} \left[\frac{\hat{M}}{2} 
     \left(\frac{d \vR{n}}{d\tau}\right)^2
   + \hat{U}_n(\vR{}) - \frac{1}{2} \Delta |\phi_n|^4\right.
 \left.
+ W\left(2\cmod{\phi_n} - \sum_{m\ne n}J_{nm}\phi_n^*\phi_m \right)
\right],
\end{eqnarray}
where $\vR{n}$ describes the position of each chain node, 
$\hat{M}$ is the node mass, $W$ is the linear excitation transfer 
energy and $\Delta$ the non-linear self-trapping interaction. 
The excitation transfer coefficients $J_{n,m}$ are of the form:
\begin{equation}\label{alpha-Eq}
J_{n,m} = J(|\vR{n}-\vR{m}|) = (e^\alpha -1)\,
           e^{-\alpha|\vR{n}-\vR{m}|/\hat{a}},
\end{equation}
where $\alpha^{-1}$ sets the relative length scale over which the interaction 
decreases, in units of $\hat{a}$, where $\hat{a}$ is the rest distance 
between two adjacent sites. 
The function $J_{n,m}$ describes the long range interaction between the 
electron field at different lattice sites $n$ and $m$;  its value decreases 
exponentially with the distance between them. Notice that when $\alpha$ is 
large and $|\vR{n}-\vR{m}|\approx\hat{a} $, 
this corresponds to a nearest neighbour 
interaction with $J_{n,m}\approx \delta_{n,m\pm 1}
(1+\alpha (1-\frac{|\vR{n}-\vR{m}|}{\hat{a}}))$.

In our formulation of the model, the normalisation of the electron field 
is preserved 
\textit{i.e.}
\begin{equation}
\sum_n |\phi_n|^2 = 1.
\end{equation}

The phonon potential $\hat{U}_n$ consists of three terms:
\begin{eqnarray}
\hat{U}_n(\vR{})= 
\frac{\hat{\sigma}}{2} (|\vR{n}-\vR{n-1}|-\hat{a})^2 +
\frac{\hat{k}}{2} \frac{\theta_n^2}
                       {\left[1-(\theta_n/\theta_{max})^2\right]}
\nonumber \\
+\frac{\hat{\delta}}{2} \sum_{m\ne n} (\hat{d}-|\vR{n}-\vR{m}|)^2
                 \Theta(\hat{d} -|\vR{n}-\vR{m}|),
\label{defUphys}
\end{eqnarray}
where the Heaviside function is defined as $\Theta(x) = 1$ for $x > 1$ and
$\Theta(x) = 0$ for $x < 1$.

The first two terms in $\hat{U}_n$ describe the elastic and the
bending energy of the chain respectively. 
$\hat{a}$ is the equilibrium separation between nodes and   
$\theta_n$ is the angle between 
$\vR{n}-\vR{n-1}$ and $\vR{n+1}-\vR{n}$.
Finally $\theta_{max}$ is the largest angle allowed between adjacent links.

The term proportional to  $\hat{\delta}$ in $\hat{U}_n$, models hard-core 
repulsion between the atoms of the chain. 
$\hat{\delta}$ should always be larger than $\hat{\sigma}$ and $\hat{d}$ 
will correspond to the minimum distance allowed between nodes.  

For convenience, the symbols denoted by an overhead carat sign \textit{e.g.}
$\hat{M}$, $\hat{\sigma}$ \textit{etc.} correspond  to physical
variables carrying units and dimensions while those without it
correspond to dimensionless variables and parameters described below,
except $H$, $\Delta$ and $W$ which are dimensional quantities. 
We also use the symbol $\vec{R}$ for position of the nodes in physical units 
and  $\vec{r}$ in dimensionless units.
First we define the time scale $\tau_0=\hbar \Delta/W^2$
and use the lattice spacing $\hat{a}$ as the length scale. We can then define
the dimensionless time $t$, position $r$ and coupling constant $g$ as 
\begin{eqnarray}
t &=& \frac{\tau}{\tau_0}
\qquad\qquad
g = \frac{\Delta}{W}
\qquad\qquad
\vr{} = \frac{\vR{}}{\hat{a}}.
\end{eqnarray}
In terms of these variables the Hamiltonian takes the form
\begin{eqnarray}\label{eqH}
H &=& \frac{W^2 }{\Delta} \sum_{n}
 \left[ \frac{M}{2}\left(\frac{d\vr{n}}{dt}\right)^2
   +  U_n(\vr{})\right.\nonumber\\
&&\left.
   +  g\left(2\cmod{\phi_n} - \sum_{m\ne n}J_{nm}\phi_n^*\phi_m \right)
            - \frac{g^2}{2} |\phi_n|^4
\right],
\end{eqnarray}
where
\begin{eqnarray}
U_n(\vr{})&=& \frac{\sigma}{2} (|\vr{n}-\vr{n-1}|-a)^2 +
\frac{k}{2} \frac{\theta_n^2}
                 {\left[1-(\theta_n/\theta_{max})^2\right]}
\nonumber\\
&&
+ \frac{\delta}{2} \sum_{m\ne n} (d-|\vr{n}-\vr{m}|)^2
                      \Theta(d -|\vr{n}-\vr{m}|)
\end{eqnarray}
with
\begin{eqnarray}
M &=& \hat{M} \frac{\hat{a}^2 W^2}{\hbar^2 \Delta}
\qquad\qquad
\sigma = \hat{\sigma}\frac{\hat{a}^2\Delta}{W^2}
\qquad\qquad
\delta = \hat{\delta}\frac{\hat{a}^2\Delta}{W^2}\nonumber\\
k &=& \hat{k} \frac{\Delta}{W^2}
\qquad\qquad
a = 1
\qquad\qquad
d = \frac{\hat{d}}{\hat{a}}.
\end{eqnarray}

Writing $\vr{n}=(x_{1,n},x_{2,n},x_{3,n},)$ we can derive the equation 
of motion for $x_{i,n}$ from the Hamiltonian (\ref{eqH}): 
\begin{eqnarray}\label{Polaron-Chain-Dynamics-Eq}
M\frac{d^2x_{i,n}}{dt^2} + \Gamma \frac{d x_{i,n}}{dt} + F(t)
  +\sum_m \frac{dU_m}{dx_{i,n}} &&
\nonumber\\
  -g\sum_k \sum_{m < k}\frac{d J_{km}}{dx_{i,n}}
               (\phi_k^*\phi_m+\phi_m^*\phi_k) &=& 0
\nonumber\\
i\frac{d\phi_n}{dt}-2 \phi_n + \sum_{m\ne n} J_{nm}\phi_m
   + g \cmod{\phi_n}\phi_n
   &=& 0,
\end{eqnarray}
where the force $F(t)$ and the friction term $\Gamma\,d x_{i,n}/dt $,
were added by hand to incorporate thermal 
fluctuations and $F(t)$ was chosen as a delta correlated white noise satisfying
\begin{equation}
<F(0) F(s)> = 2 \Gamma k_{B} T \delta(s)
\end{equation}
where
\begin{equation}
k_{B} T = \hat{k}_B \hat{T} \frac{\hat{W}^2}{\hat{\Delta}}
         = \hat{k}_B\hat{T} \hat{W} g.
\end{equation}
As the equation for $x_i$ is expressed in units of 
$\hat{W}^2/(\hat{\Delta} \hat{a})$,
we have $\Gamma = \hat{\Gamma} \hat{a}^2/\hbar$. 
The friction coefficient $\hat{\Gamma}$ can be evaluated from  
$\hat{\Gamma}\approx6\pi\mu R_0$ where $\mu= 0.001 \mbox{Pa\, s}$ 
for water, and up to 4 times that value for the cytoplasm, 
where $R_0$ is the average radius of a 
single molecule of the lattice. 
Notice also that the electron field $\phi_n$ is coupled
to the phonon field $x_{i,n}$ through the function $J_{nm}$.

In what follows we are primarily interested in stationary
configurations. To compute such solutions numerically we choose an
initial lattice configuration with a loop structure and localised 
the electron so that it overlapped with both tails of the loop.
We achieved this by using an approximation for the polaron electron field and 
distributing it over a few nodes spread between the two ends of chain.
This way the polaron was able to bind the loop extremities together.
We then relax the electron field keeping the lattice
configuration unchanged and then evolved the entire system with an 
absorption term until it relaxed to a static configuration. 
This was achieved by solving equation 
(\ref{Polaron-Chain-Dynamics-Eq}) without thermal noise. 

In all our simulations we started from a very small value of $k$, typically
$k=0.005$, so that the lattice offered very little resistance to bending. 
We then increased the value of $k$ in small increments using
the relaxed conformation obtained for the previous $k$
value as the initial configuration. We then equilibrated the system
for the new value of $k$. By repeating the procedure for each value
of $g$ we have determined the critical value $k_{crit}(g)$ up to which the
given configuration can be sustained by the polaron. 

Unless otherwise stated, we have used the following parameter values: 
$\delta=10000$, $\sigma=1000$ and $M=0.5$. For stationary solutions the mass 
term does not affect the results and $\delta$ was chosen so that the repulsion
potential is close to that of a hard shell. Finally for all the  
computed configurations, nothing prevents the nodes from being very close 
to their equilibrium 
distance and hence we have selected a relatively large value for $\sigma$ to
approximate stiff cross-node links.
Following Mingaleev et al, we have also 
considered mainly the case $\alpha=2$ and $d=0.6$. Finally we have also 
considered the effect of varying the values of these two parameters.

To solve equation (\ref{Polaron-Chain-Dynamics-Eq}) we used a 4th 
order Runge-Kutta
method with a time step $dt=0.0001$ in dimensionless units. To compute static 
configurations, we took  $T=0$, {\it i.e} no thermal noise, setting 
$\Gamma=1$ and
then integrating equation (\ref{Polaron-Chain-Dynamics-Eq}) 
until the system relaxed to a stationary solution.

To study the thermal stability of the configurations for DNA and polyacetylene,
we solved equation
(\ref{Polaron-Chain-Dynamics-Eq}), taking $T=300K$ and estimated 
$\Gamma$ from the radius of the molecules as described above. For those 
simulations
we started from the static configuration for which we wanted to evaluate the 
stability and let the system thermalise itself. The time needed for this 
thermalisation was always orders of magnitude smaller than the average life time
of the configurations we considered and so we did not need to resort to 
sophisticated thermalisation procedure as we did in \cite{CPZ12}.

\section{Plain Loop Configurations}
\label{Sec:Loop}
\begin{figure}[ht]
\unitlength1cm \hfil
\begin{picture}(14,4)
\epsfxsize=4cm \epsffile{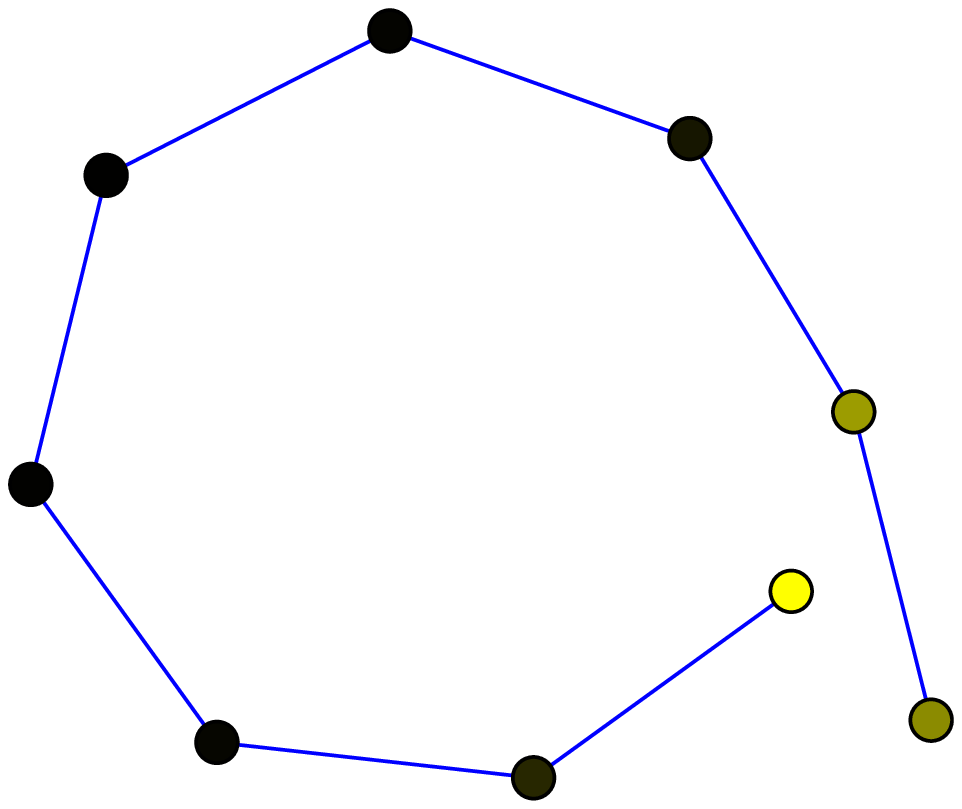}
\epsfxsize=4cm \epsffile{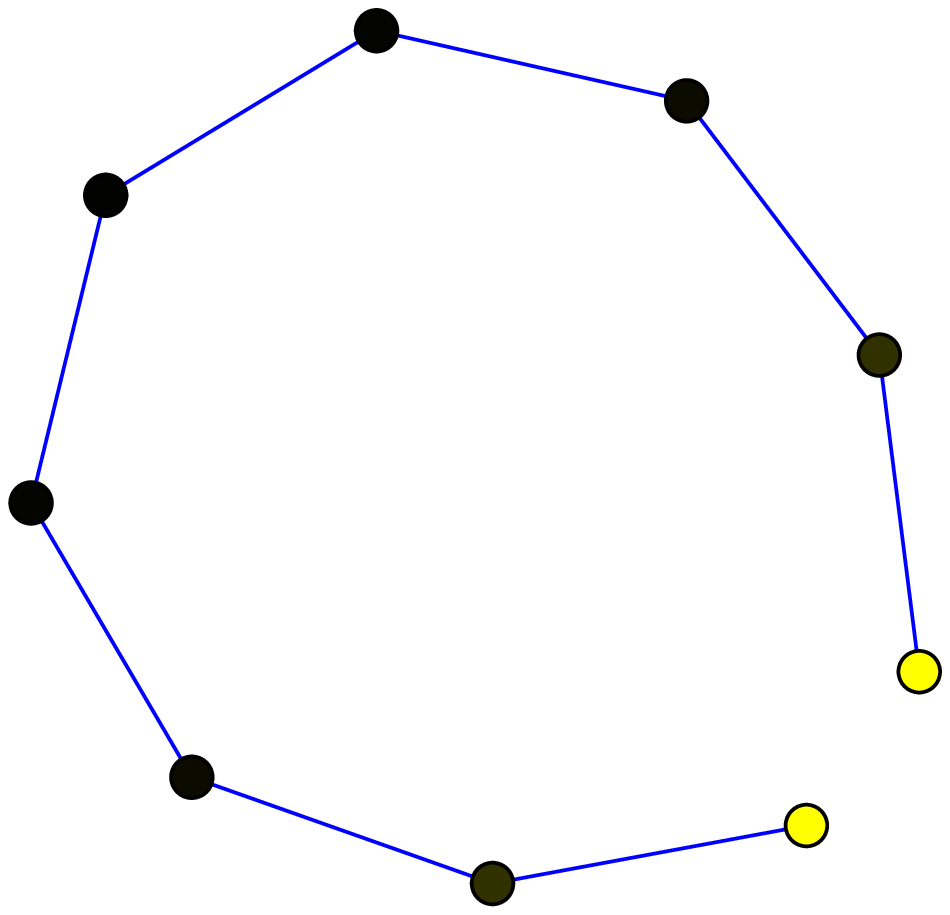}
\epsfxsize=4cm \epsffile{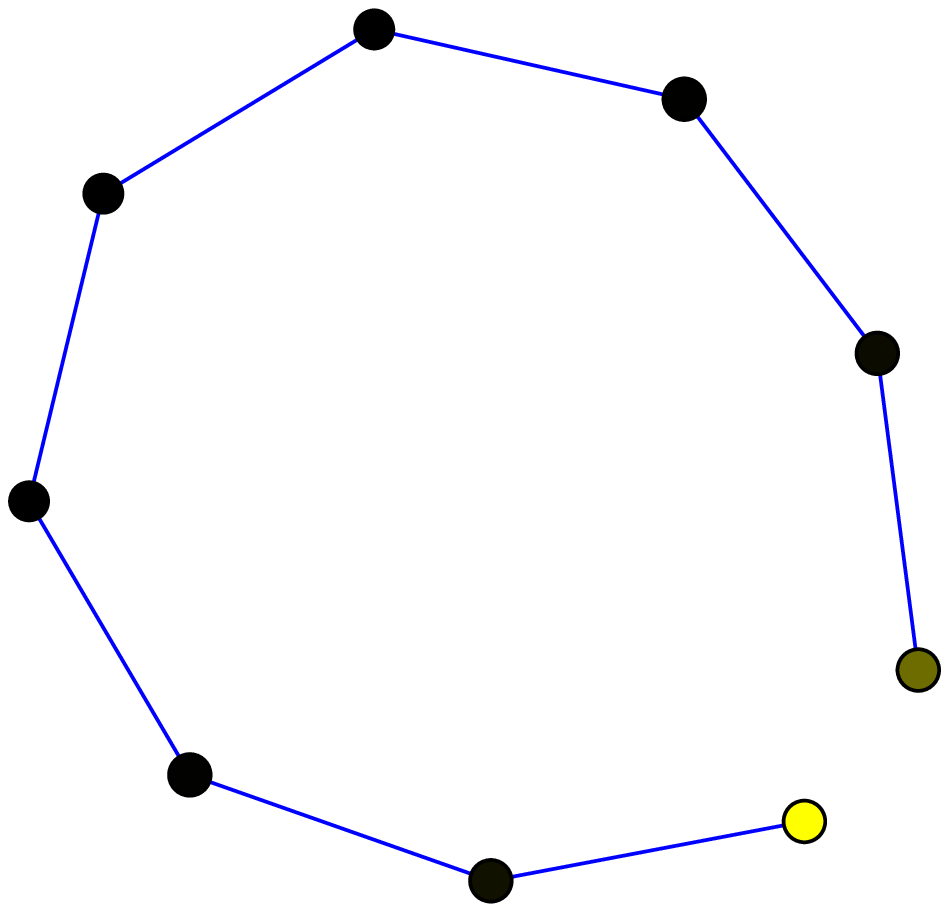}
\end{picture}\break
\begin{picture}(14,0.2)
\put(2,0){(a)}
\put(6,0){(b)}
\put(10,0){(c)}
\end{picture}\break
\caption{Loop configuration for $N=9$ nodes for
a) $g=2.5$, $k=0.5$, $|\phi_0|^2=0.407$, $|\phi_8|^2=0.221$, 
   $|\phi_7|^2=0.249$; 
b) $g=2.5$ $k=4$, $|\phi_0|^2= |\phi_8|^2=0.397$; 
c) $g=5$ $k=5$, $|\phi_0|^2=0.642$, $|\phi_8|^2=0.272$.}
\label{Fig1}
\end{figure}

Our first investigation involved considering a simple loop configurations for 
which all the nodes lie 
more or less on a circle with the two end points close to each other 
(separated by a distance $d$). When $k$ is very small, the favoured 
configuration is one similar to the one presented in figure \ref{Fig1}.a.
In this figures, the electron probability density is represented by the colour
of the node. A dark colour corresponds to a null value while a light value
corresponds to a higher probability density.
The node at which the polaron field has its maximum value is close to the last
2 points at the opposite ends of the chain. 
This allows the electron field to be distributed on 2 nearby nodes
rather than a single one and, as $k$ is small, the deformation of the 
chain does not prevent this from taking place. 
As $k$ increases, such a localisation becomes energetically expensive and the
configuration assumes the shape of a horse-shoe, as presented in 
figures \ref{Fig1}.b. and \ref{Fig1}.c. As one increases $k$ further, 
there is a point at which the stretching energy is too large and the polaron is 
not able to sustain the loop anymore. 

The difference between figures \ref{Fig1}.b. and \ref{Fig1}.c is that in 
the former
the electron is localised equally on the 2 end nodes while in the later it is
localised mostly on a single node. The difference is dictated by the value of 
$g$: for small $g$, the polaron is wide and the electron spreads 
itself nearly equally between the two end points of the chain 
(Fig \ref{Fig1}.b). 
As $g$ increases, the polaron becomes more localised and the electron 
becomes localised, more asymmetrically, on a single node (Fig \ref{Fig1}.c).

\begin{figure}[ht]
\unitlength1cm \hfil
\begin{picture}(14,4)
\epsfxsize=4cm \epsffile{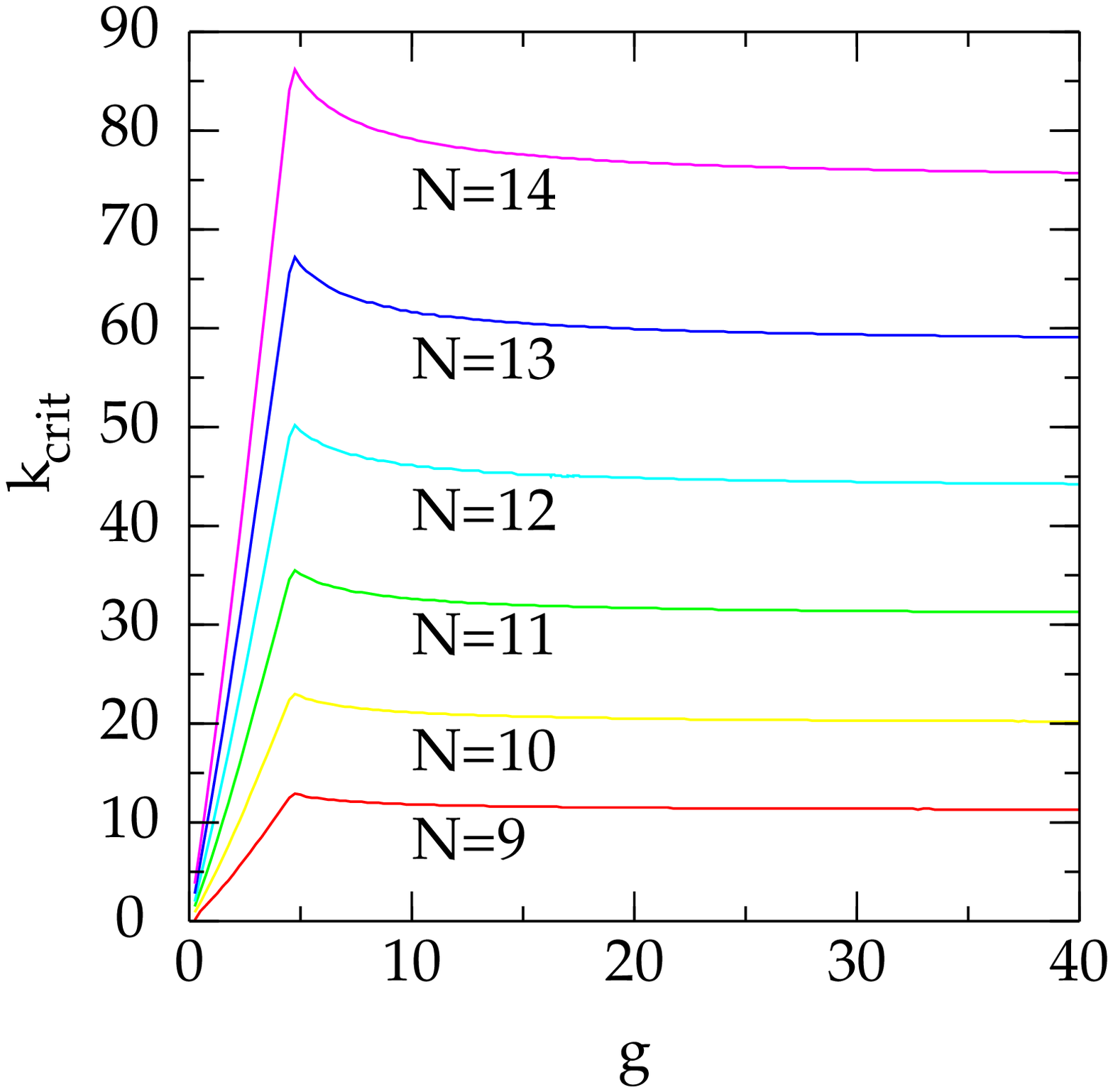}
\epsfxsize=4cm \epsffile{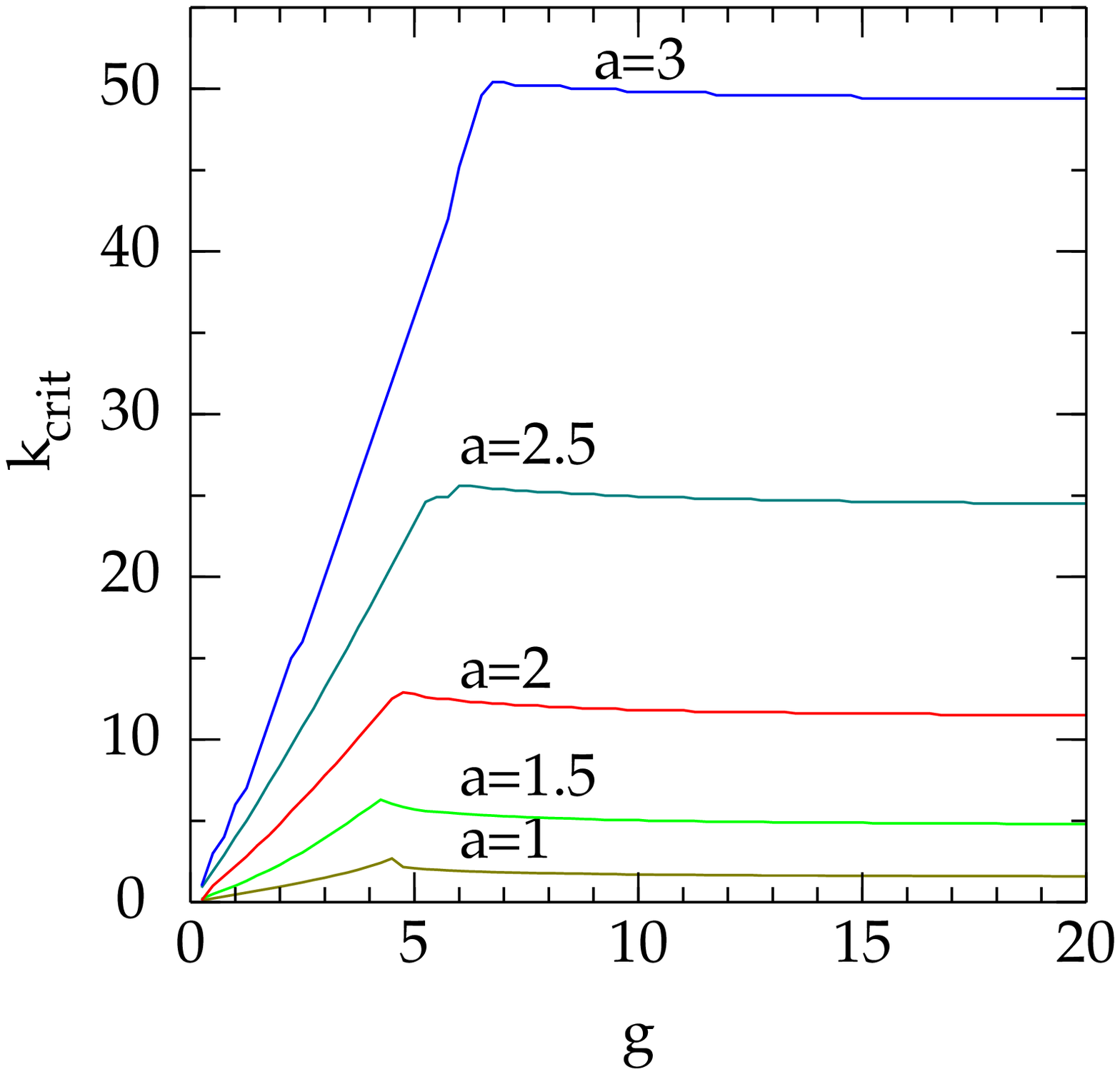}
\epsfxsize=4cm \epsffile{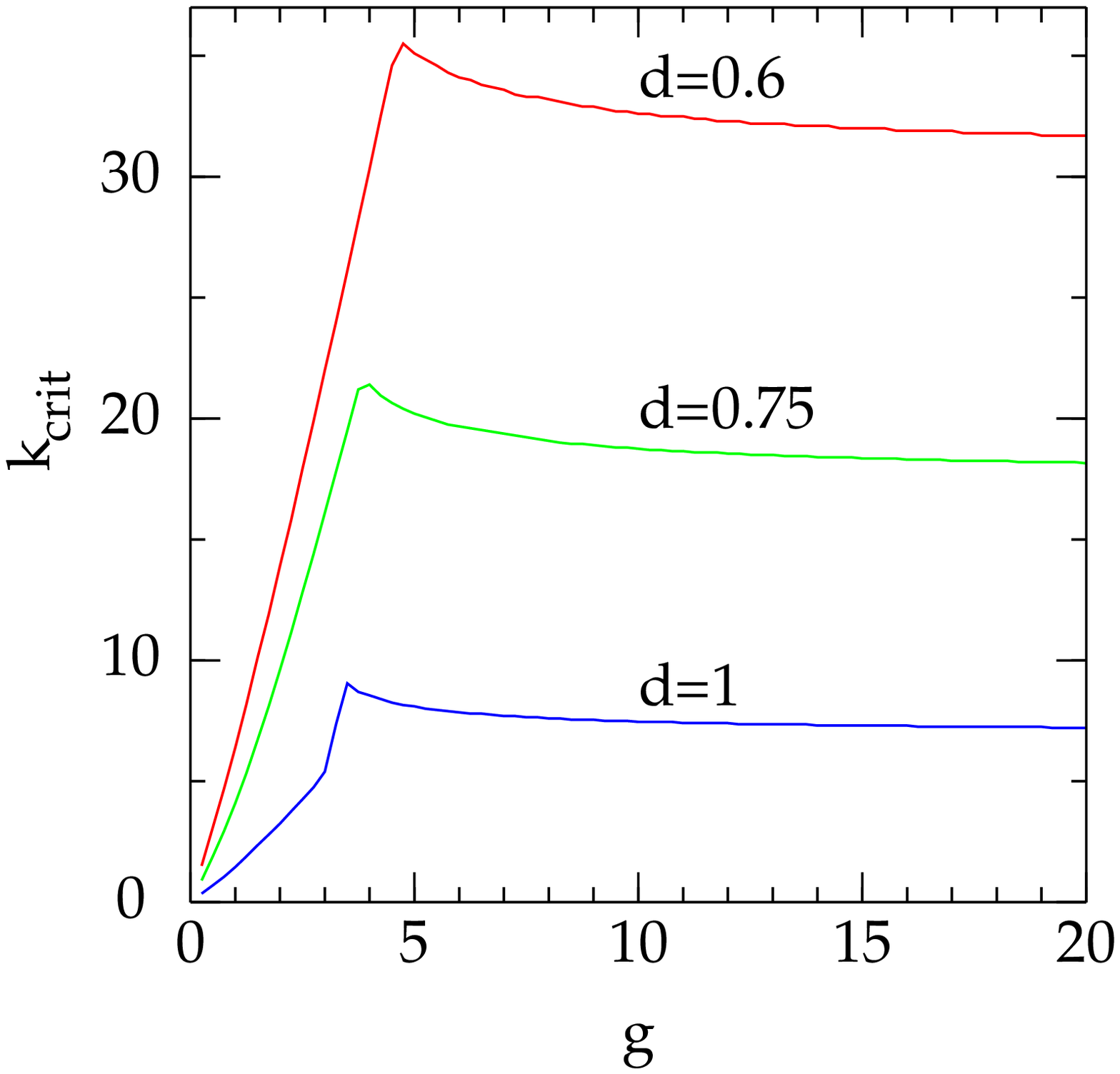}
\end{picture}\hfill
\begin{picture}(14,0.2)
\put(2,0){(a)}
\put(6,0){(b)}
\put(10,0){(c)}
\end{picture}\break
\caption{Critical value of $k$ for the existence of a loop configuration.
a) $\alpha=2$ and $N=9$ to $14$ nodes. 
b) $N=9$ nodes $\alpha=1,1.5,2,25,3$ and $d=0.6$.}
c) $N=11$ nodes $d=0.6, 0.75, 1$ and $\alpha=2$.
\label{Fig2}
\end{figure}

The critical value of $k$ as a function of $g$ is presented in 
Figure \ref{Fig2}.a for loops consisting of 9 to 14 nodes.
It is interesting to note that when $g$ is small, the critical value of $k$ 
is small.
This can be explained by the fact that the coupling parameter $g$ is small
but also by the fact that the polaron is delocalised and hence
the fraction of the electron close to the end point is smaller than for larger 
values of $g$. The maximum value of $k_{crit}$ is reached for $g\approx 5$.
For very large values of $g$, the electron is nearly fully 
localised on a single lattice point, but the attraction exerted by the polaron,
surprisingly decreases very slowly.

Having followed \cite{Mingaleev02} and taken the values $\alpha=2$ and 
$d=0.6$ for the results presented so far, it is worth checking how these two 
parameters affect the results that we have obtained. We started by varying 
$\alpha$, which controls, through $J_{n,m}$, how fast the coupling between 
nodes decreases with the distance separating them. 
The results are presented in Figure 
\ref{Fig2}.b where we see that $k_{crit}$, 
contrary to what one might expect,  increases with $\alpha$. This is easily 
explained: having chosen $d=0.6$, increasing $\alpha$, not only reduces the
long distance interaction between nodes but it also increases exponentially 
the binding energy of nodes that are very close to each other.
The binding energy of the end nodes, which are separated by a 
distance $d < 1$, thus increases with $\alpha$. For this reason, we
have decided to take $d=a=1$ when we consider single stranded DNA and 
polyacetylene later in the paper.  

In figure \ref{Fig2}.c we show how the critical value of $k$ varies with $d$.
As the parameter $d$ sets the minimal distance allowed between 2 nodes and 
given that
$J_{n,m}$ decreases with the distance between nodes, it is not surprising 
that $k_{crit}$ decreases when $d$ becomes larger, but loop 
configurations can still be held by the polaron.

\section{Hairpin-Loop Configurations}
\label{Sec:Hairpin}
\begin{figure}[ht]
\unitlength1cm \hfil
\begin{picture}(14,4)
 \epsfxsize=7cm \epsffile{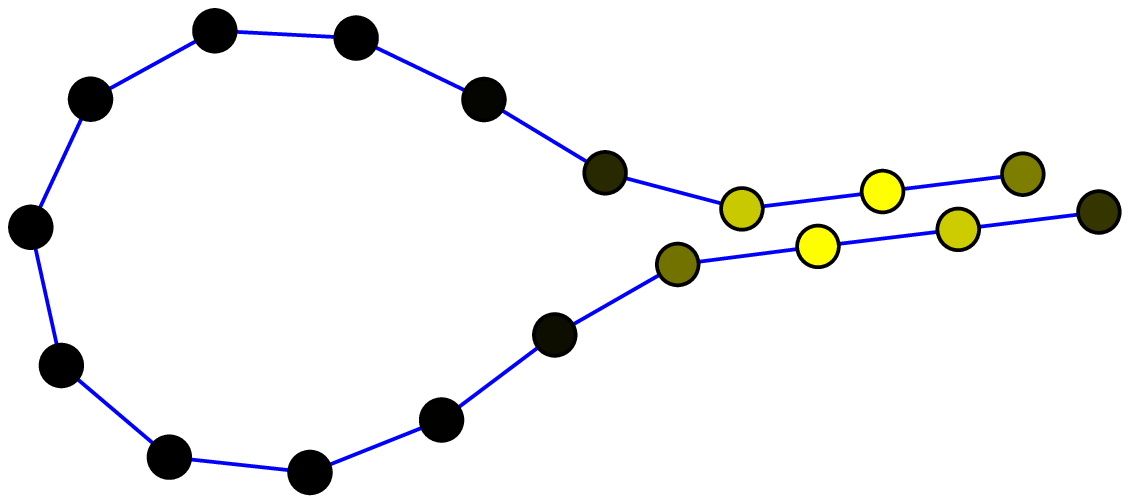}
 \epsfxsize=7cm \epsffile{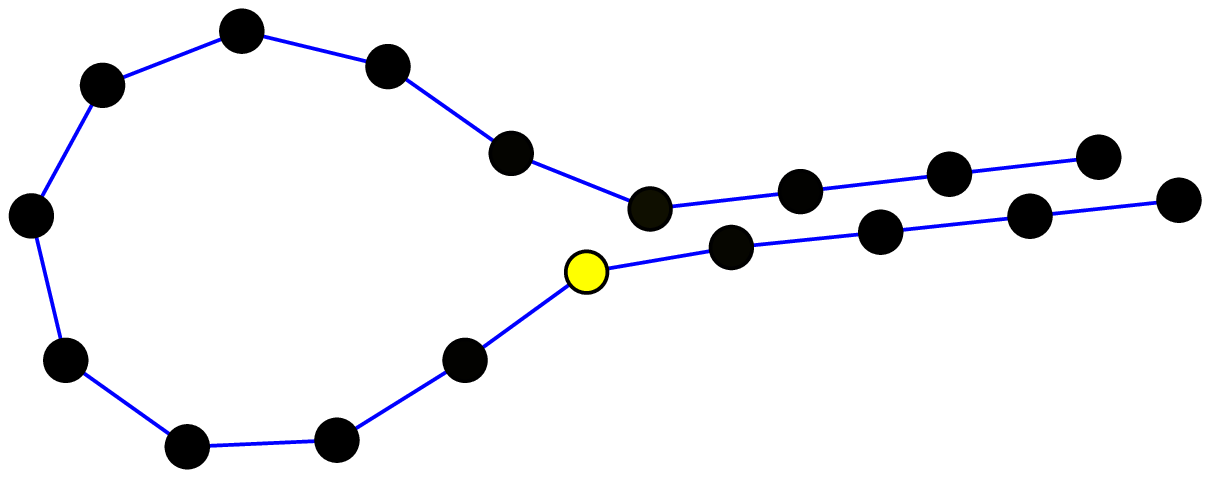}
\end{picture}
\begin{picture}(14,0.2)
\put(2,0){(a)}
\put(8,0){(b)}
\end{picture}\break
\caption{Hairpin-loop configuration for $N=18$ nodes. 
The brightness of the nodes is proportional to $|\phi|^2$.
a) $g=1.5$ and $k=0.5$  (max $|\phi|^2= 0.188$), 
b) $g=11$ and $k=10$ (max $|\phi|^2=0.887$ )}
\label{Fig3}
\end{figure}

Now we consider a hairpin-loop configuration as presented in Figure
\ref{Fig3} similar to the structure that single stranded DNA can form and which
is potentially more relevant to long chains. As for the plain loops, we 
generated these configurations for a small $k$ and then 
slowly increased its value until the number of links, $L$, making the loop 
increased by one unit. This gave us the critical value, $k_{crit}$, 
for which the hairpin-loop configuration of a given size can be sustained by 
the polaron.

\begin{figure}[ht]
\unitlength1cm 
\begin{picture}(14,5)
  \epsfxsize=6cm \epsffile{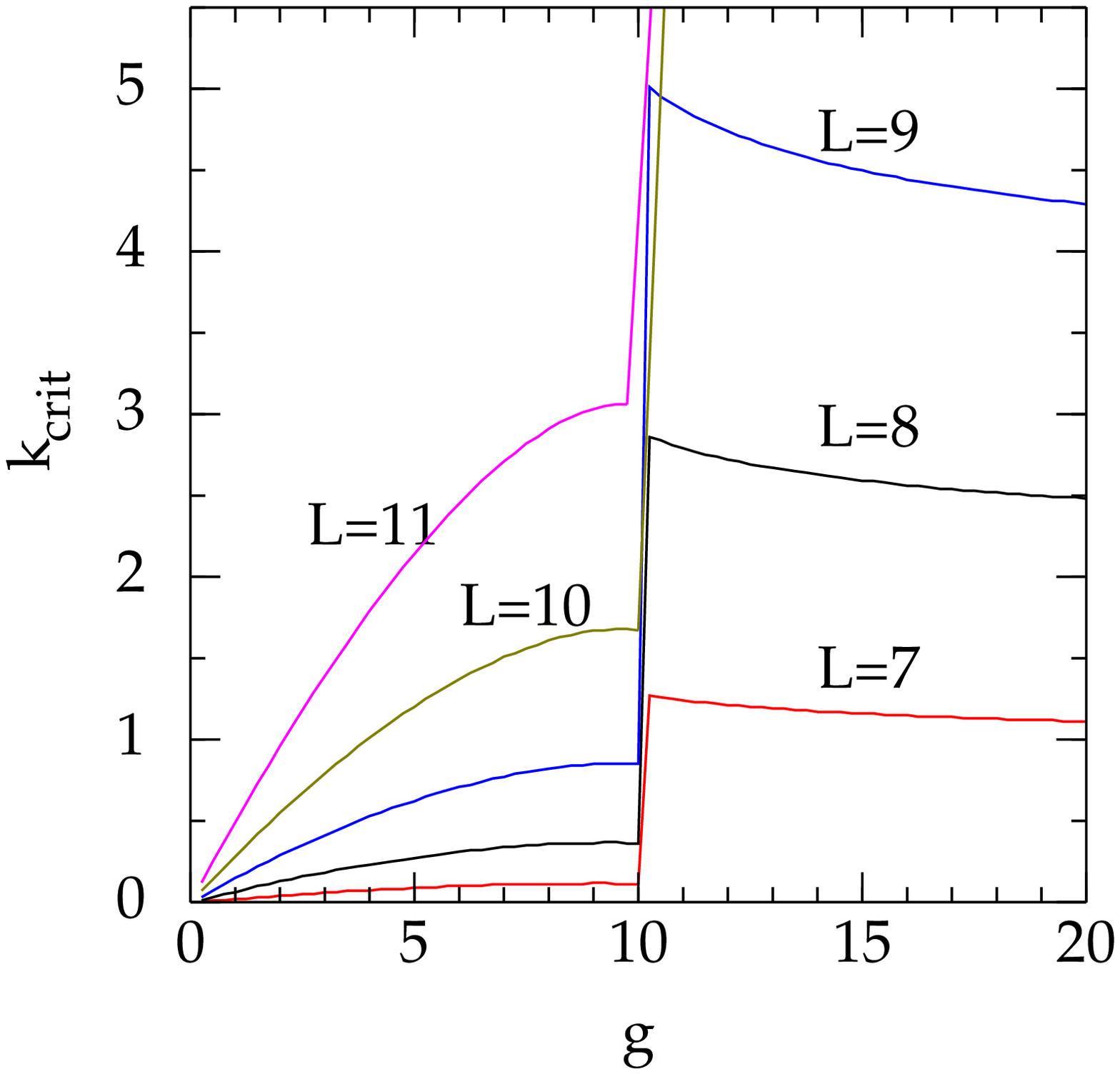}
  \epsfxsize=6cm \epsffile{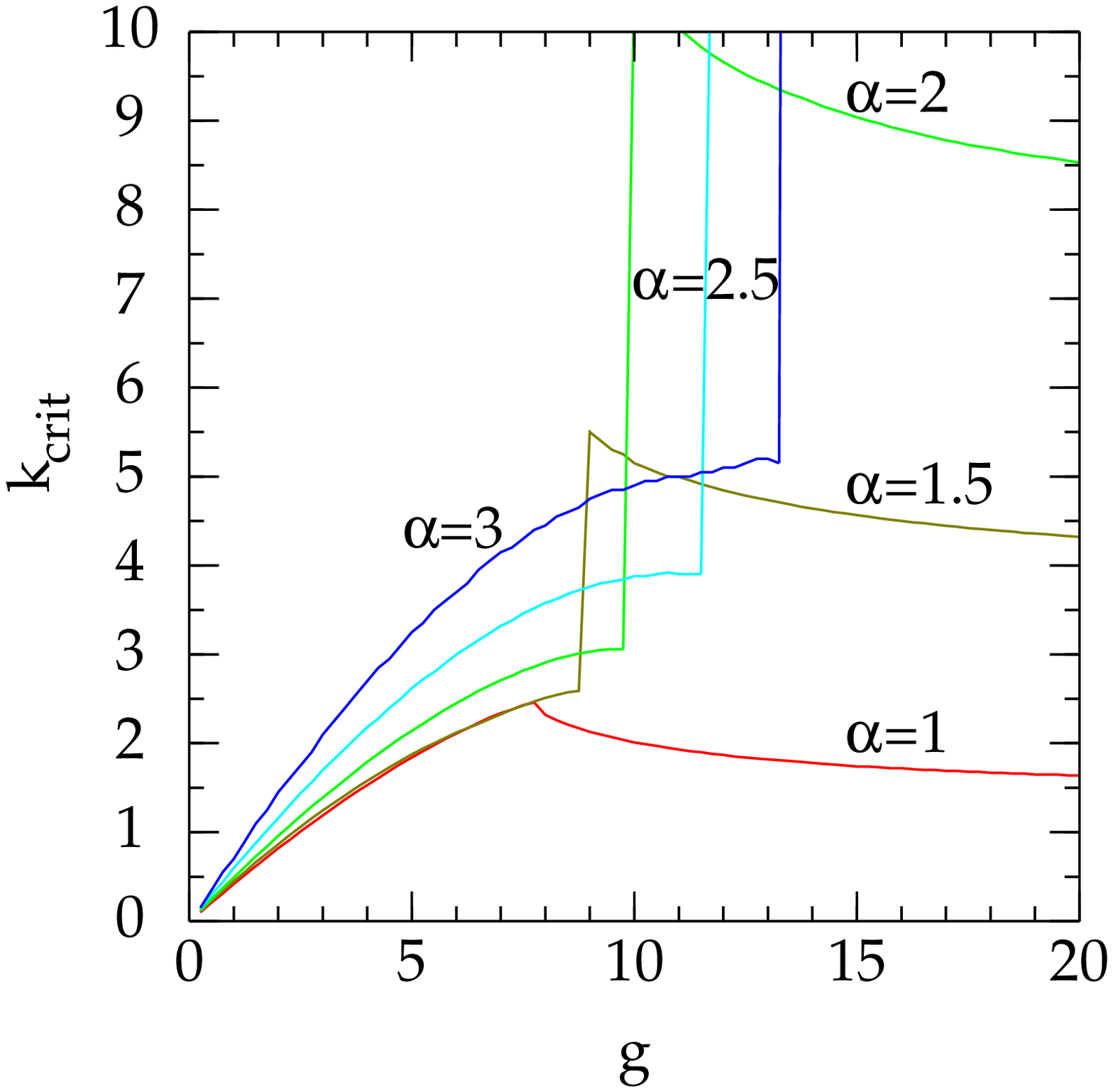}
\end{picture}
\begin{picture}(14,0.2)
\put(3,0){(a)}
\put(9,0){(b)}
\end{picture}
\hfill\break
\caption{Critical value of $k$ for the existence of an hairpin-loop 
configuration.
a) $\alpha=2$ and $N=7$ to $11$ node loops. 
b) $N=9$ nodes and $\alpha=1,1.5,2,2.5,3$.}
\label{Fig4}
\end{figure}

The results are presented in figure \ref{Fig4} where we can see a
sharp transition around $g=10$. Below this value the hairpin-loop is
only viable for relatively small values of $k$ but above it, they are
sustainable for much more rigid chains. This is due to the fact that for
small values of $g$, the polaron is always distributed over the handle of the 
hairpin-loop while when $g>10$, it is localised mostly on one lattice site,
at the base of the loop. In that case the interaction is stronger and supports
loops for larger values of $k$.

To make sure this was not an artefact of our procedure, we have tried to 
construct solutions using various initial conditions. We also used solutions
obtained for $g>10$ as initial conditions and then slowly decreased the value 
of $g$. Regardless of the procedure we used, we always obtained the curve 
of Figure \ref{Fig3}.a. 

As expected, the configurations of figure \ref{Fig3} 
are harder to sustain than a simple loop as
the chain needs to be bent near the stem of the hairpin-loop. 

\section{Analytic Approximation}
\label{Sec:Results-Analytical}
\begin{figure}[ht]
\unitlength1cm \centerline{
\begin{picture}(5,5)
 \epsfxsize=5cm
  \epsffile{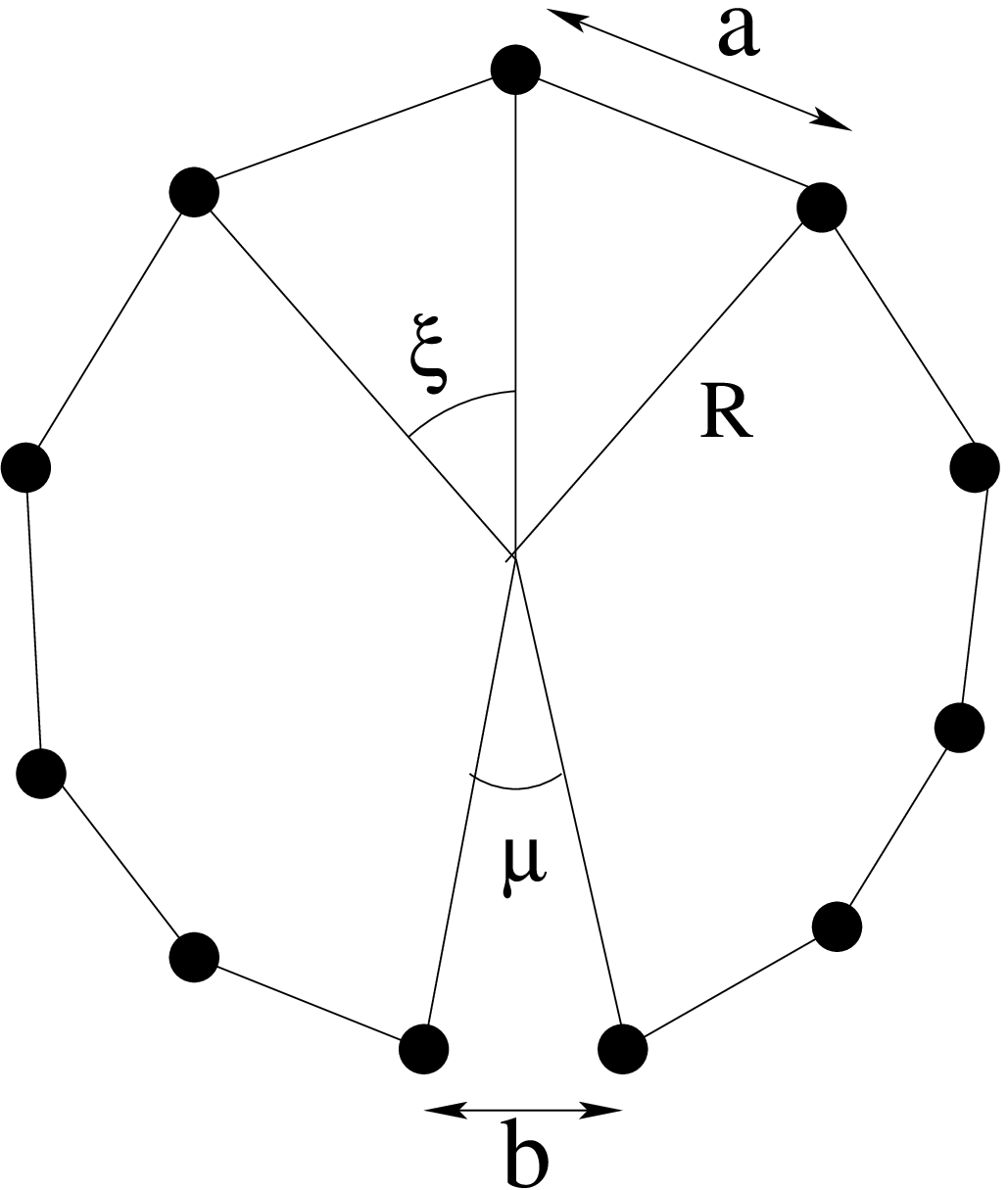}
\end{picture}}\break
\caption{Schematic representation of a polymer-loop with $N$ nodes.
$N-1$ bonds with rest distance $a$ subtend an angle $\xi$ at the centre. 
The two end nodes, on which the electron is localised, are separated by 
a distance $b$ and span an angle $\mu$ at the centre.
}
\label{Fig5}
\end{figure}

Having computed numerically the critical value $k_{crit}(g)$ for which the 
polaron is able to sustain a loop of a given size, we  
now try to estimate this value analytically. To do this, 
we consider a circular 
configuration of radius $R$ made out of $N$ segments, 
one of length $b$ and $N-1$ of length $a$, as depicted in figure \ref{Fig5}.
Note that the two nodes separated by the distance $b$ are not linked to 
each other.
If $\xi$ and $\mu$ are the angles opposite $a$ and $b$, respectively, we
have
\begin{eqnarray}
\begin{array}{lll}
(n-1)\xi + \mu = 2 \pi\qquad 
&
\sin(\frac{\xi}{2}) = \frac{a}{ 2R}\qquad 
&
\sin(\frac{\mu}{2}) = \frac{b}{ 2R}
\end{array}
\end{eqnarray} 
and so
\begin{eqnarray}
\sin(\frac{\xi}{2}) = \frac{a}{b} \sin(\frac{\mu}{2}) 
                    = \frac{a}{b} \sin(\frac{(n-1)\xi}{2}).
\label{eq_bxi}
\end{eqnarray} 

Choosing specific values $a=1$,  $b=b_0$, this transcendental equation can be 
solved numerically to obtain the corresponding value $\xi=\xi_0$.
We can then perform a first 
order expansions around this solution: 
$b = b_0 +\delta b, \xi = \xi_0+\delta \xi$ and obtain
\begin{equation}
\delta\xi = \frac{ \sin(\frac{\xi}{2})}
{\frac{N-1}{4}\cos((N-1)\frac{\xi}{2})-\frac{b_0}{4}\cos(\frac{\xi}{2})}
\delta b
\label{dbdxi}
\end{equation}

To determine the critical value of $g$ and $k$ for which a loop configuration 
can exist, we have to minimise the Hamiltonian and, for each value of $g$ 
and $b$, 
determine the value of $k$ for which this Hamiltonian has a minimum. 
For each $g$, we 
then select the value of $b$ for which $k$ is the largest.

Let us assume that the loop is symmetric, so that 
the $N-2$ bending terms are all 
identical and are functions of $\xi$. The elastic terms are then also equal,
but as they do not depend on $\xi$, they are constant and can thus 
be ignored for the minimisation. 
The repulsion term proportional to $\delta$ can also be ignored if $b> d$.
When $b<d$, the repulsion term leads to a very large energy 
increase and we can thus consider than $b=d$ is the smallest value we should 
consider.

To evaluate the electron field, we take the continuum limit of
equation \ref{Polaron-Chain-Dynamics-Eq} for stationary solutions: 
\begin{equation}
\frac{ d^2 \phi_c}{dx^2} + \hat{g} |\phi_c|^2\phi_c -\lambda \phi_c =0,
\end{equation}
which is the well known non-linear Schoedinger equation, 
where $\hat{g} = g/(1-e^{-\alpha})$, and which admits the following
solution:
\begin{equation}
\phi_c(x) = \sqrt{\frac{\hat{g}}{8}}{\cosh(\frac{\hat{g}x}{4})}.
\label{eq_phic}
\end{equation}
Note that $\int|\phi(x)|^2 dx = 1$ and  $\lambda = -\hat{g}^2/16$.
From the numerical solutions, we know that the wave function is centred
on one of the two end lattice points and we can thus 
take $\phi_0=\phi_c(0)$ and $\phi_{N-1}=\phi_c(b_0)$.

The Hamiltonian can then be approximated by the following function of $b_0$
and $\xi_0$
\begin{eqnarray}
H &\approx& -2\,g\,(e^\alpha-1)e^{-\alpha b_0/a}\phi_0\phi_{N-1} 
+\frac{k}{2} \frac{(N-2)\xi_0^2}
           {\left[1-\left(\frac{\xi_0}{\theta_{max}}\right)^2\right]}\,.
\end{eqnarray}

Next we compute the variation of $H$ with respect to $b$ and $\xi$ 
\begin{eqnarray} 
\delta H =
2 g \alpha \phi_0\phi_{N-1} (e^\alpha-1)e^{-\alpha b_0}\delta b
+ \frac{k (N-2) \xi_0}
  {\left[1-\left(\frac{\xi_0}{\theta_{max}}\right)^2\right]}
  (1-\xi_0^2(1-\frac{1}{\theta_{max}^2}))\delta \xi
\end{eqnarray}
Using equation (\ref{dbdxi}) and imposing $\delta H =0$ we get the 
condition
\begin{eqnarray} 
k_{crit} &=& \frac{g}{D} \phi_0\phi_{N-1}
\label{eq_kcrit}
\end{eqnarray}
where
\begin{eqnarray} 
D &=& 
\frac{1}
{2\alpha (e^\alpha-1)e^{-\alpha b_0}}
  \frac{(N-2) \xi_0}
  {\left[1-\left(\frac{\xi_0}{\theta_{max}}\right)^2\right]}
  (1-\xi_0^2(1-\frac{1}{\theta_{max}^2}))\nonumber\\
&& \frac{ -4\sin(\frac{\xi_0}{2})}
{(N-1)\cos((N-1)\frac{\xi_0}{2})-b_0\cos(\frac{\xi_0}{2})}
\label{eq_D}
\end{eqnarray}
which depends on $g$ but not on $k$. From equations (\ref{eq_kcrit}) 
and (\ref{eq_D}) it is 
clear that $k_{crit}$ increases as $b_0$ decreases and so we have to choose 
the smallest possible value for $b_0$ {\it i.e.} $b_0=d$.

\begin{figure}[ht]
\unitlength1cm \hfil
\begin{picture}(14,7)
  \epsfxsize=6cm \epsffile{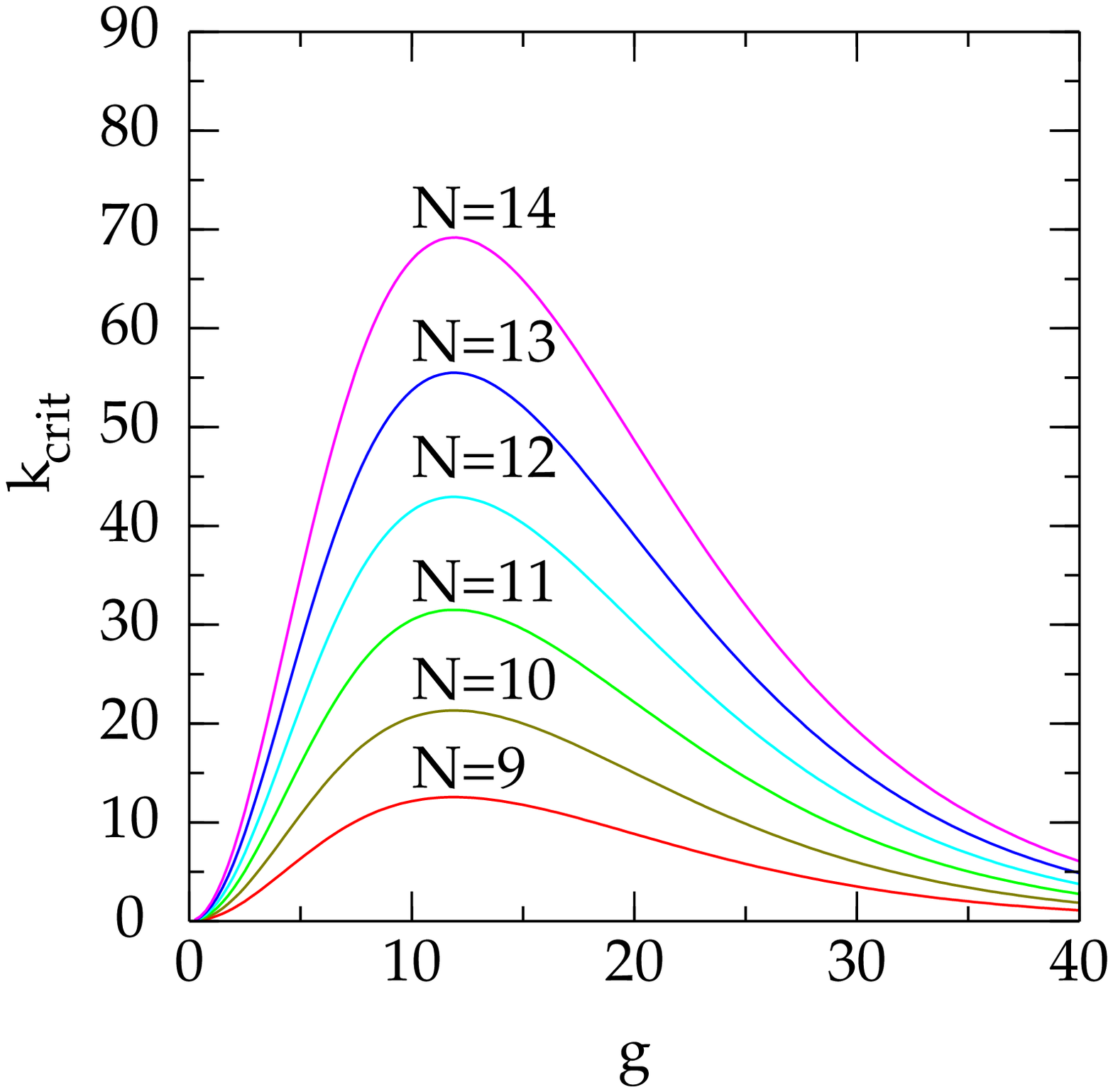}
  \epsfxsize=6cm \epsffile{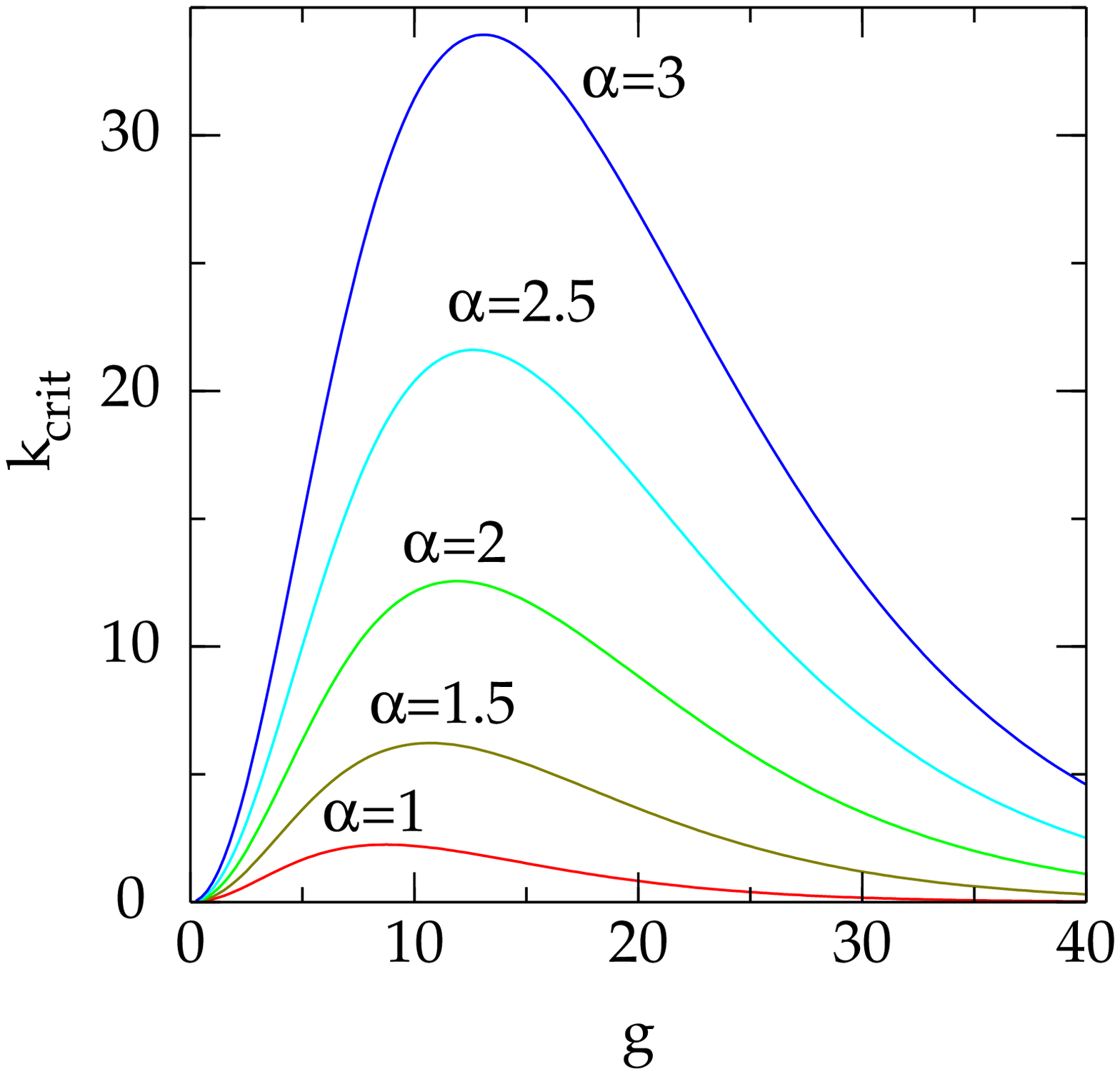}
\end{picture}\break
\begin{picture}(14,0.2)
\put(3,0){(a)}
\put(9,0){(b)}
\end{picture}
\caption{Theoretical estimation of the critical value of $k$ for the 
existence of a loop configuration as a function of $\alpha$ and $N$.
a) $\alpha=2$ and $N=9$ to $14$ nodes. 
b) $N=9$ nodes and $\alpha=1,1.5,2,25,3$}
\label{Fig6}
\end{figure}

\begin{table}
\begin{center}
\begin{tabular}{l|l|l|l|l|l|l}[ht]
N     & 9 & 10 & 11 & 12 & 13& 14\\
\hline
$\xi_0$ &0.731341&0.654972&0.593071&0.541875&0.498824&0.462115
\end{tabular}
\caption{Value of $\xi_0$ for various number of nodes $N$.}
\end{center}
\end{table}

Taking $a=1$, $b_0=d=0.6$, we can solve equation (\ref{eq_bxi}) to obtain 
the values $\xi_0$ listed in Table 1. 
which we can use to estimate  $k_{crit}$. The results are presented 
in figure \ref{Fig6} from which we see that our 
evaluation reproduces the gross features of the results obtained numerically 
(figure \ref{Fig4}):
$k_{crit}$ is small when $g$ is very small, and increases with $g$ until a 
maximum is reached. Then, as $g$ increases further $k_{crit}$ slowly 
decreases. The maximum value obtained for $k_{crit}$ is slightly smaller
than the numerical value obtained before. The biggest discrepancy 
between the numerical and
analytic results are for large $g$, but this is to be expected as this
is the limit where the polaron is strongly localised and so is less well  
approximated by equation (\ref{eq_phic}).

\section{Single Stranded DNA}
\label{Sec:ssDNA}
Having considered the Mingaleev et al. polaron model in general, we now 
consider two explicit cases: DNA and polyacetylene, both of which have 
parameters allowing the polaron to sustain loops.

The parameters of our model were obtained from several sources.
First of all, $\hat{k}$ can be determined from the flexural rigidity,
$\hat{k} = \lambda \hat{k}_B\hat{T}/R_0$ where $R_0$ is the radius of the 
DNA strand.
We do not have experimental values of $\hat{\delta}$, but its actual value does
not play an important role except that it must be large enough to mimic
a hard shell repulsion. In practice, we chose a value larger than 
$\hat{\sigma}$.

For single stranded DNA we have $R_0\approx0.33 nm$~\cite{Mandelkern}
$\Delta\approx 0.4eV$, $W\approx 0.3eV$~\cite{Mingaleev02}, 
$\hat{k}\approx 0.11eV$~\cite{Mandelkern} and  
$\hat{\sigma}\approx 1.5eV/A^2$~\cite{Smith96}.
This leads to the following dimensionless values:
\begin{eqnarray}
\begin{array}{llll}
g \approx 1.33 &\sigma \approx 72.51 & k\approx0.487 & M\approx2.5\times 10^5
\nonumber\\
k_BT =  \hat{k_BT} \frac{\hat{\Delta}}{\hat{W}^2}\approx 0.115 & 
\Gamma \approx 3210 &
\tau_0\approx 2.92\times 10^{-15}s &
\end{array}
\end{eqnarray}

We thus see that single stranded DNA sits at the bottom left region of figures
\ref{Fig2}, \ref{Fig4} and \ref{Fig6}. For our simulations, we have chosen 
$\alpha=2$, $\delta=100$ and $d=1$, the later parameter was taken as the worst
case we could consider. We then found that DNA can easily sustain loops of $10$ 
segments and hairpin-loops with $11$ segments. 
We then studied the thermal stability of the configurations that we 
have obtained at $T=300K$. To achieve this, we started from a static 
configuration that we had obtained for DNA. 
We then solved equation (\ref{Polaron-Chain-Dynamics-Eq}), 
including the thermal 
noise, and ran 100 simulations for an extended period of time.   
We started by running 100 simulations for a loop made out of $10$ nodes 
and we measured an average unfolding time of $1ns$.
We also observed that a chain made out of $11$ nodes is much more stable, 
and experiences a very slow unfolding of the loops with an average decay time 
of approximately $1.3\mu s$.

We have also performed thermal simulations for an hairpin-loop configuration 
of $18$ nodes and $L=11$ and did not observe a single
unfolding of the chain in $20\mu s$. As the integration time steps was 
approximately
$3\, 10^{-17}s$ this required $100$ simultaneous simulations each performing
around $10^{10}$ integration steps and we decided not to run it any longer as 
the stability of the loop was sufficiently well established.
 
\begin{figure}[ht]
\unitlength1cm \hfil
\begin{picture}(14,7)
 \epsfxsize=7cm \epsffile{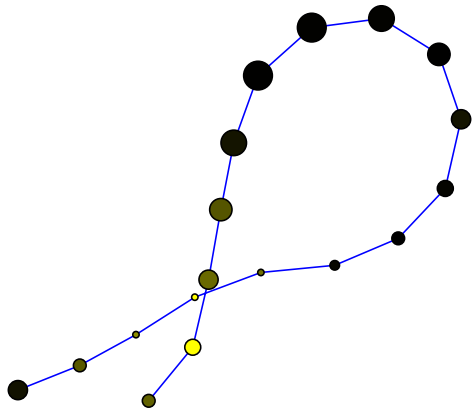}
\end{picture}
\caption{Thermalised, $T=300K$, hairpin-loop DNA configuration for $N=18$ nodes.
The size of the disk are an exaggerated indication of their depth in the 
direction transverse to the plane of view. (The two nodes close to the 
crossing point are separated by the same distance as two neighbour points). 
The brightness of the nodes is proportional to $|\phi|^2$.}
\label{Fig7}
\end{figure}

Under thermal noise, the stem,  made out 
of the the two parallel ends of the chain, deforms itself and the chain 
takes the shape of a loop where the polaron 
links the two opposite ends of the chain around a couple of nodes,
as presented in figure \ref{Fig7}.
In our simulations we observed that as the polaron moves along the chain, 
the size of the loop that it formed
fluctuated constantly in time but it never unfolded. 
We can thus conclude that the DNA 
polarons loops are very stable.

This loop configuration could play an important role in the formation of
single stranded DNA hairpin-loops in vivo. 
The formation of such configurations depends
on the likelihood of complementary DNA bases to face each other before
they can bind by hydrogen bonding and this likelihood decreases rapidly as the 
length of the chain increases \cite{Bonnet1998} and 
partially matching DNA base sequences are less stable than
perfectly matched ones~\cite{Kenward2009}. While homogeneous
sequences of DNA bases can bind quite rapidly like the one used in
\cite{Bonnet1998} and \cite{Kenward2009} for example, irregular sequences, 
like $ATGCAGTC ... GACTGCAT$ are less likely to match purely randomly.
As the polaron folds the ssDNA into a loop and moves along the chain,
the DNA bases on the opposite segments of the loop slide relative to each 
other, under the action of thermal excitation, 
increasing the probability for a complementary 
sequence of bases to face each other and bind. In vitro, the reaction time 
of hairpin-loops has been determined to be several $\mu s$~\cite{Bonnet1998}, 
a length of time that, as we have shown, DNA-polaron can outlive easily.  
Hence, we can conclude that DNA polarons can increase 
the rate of formation of hairpin-loops. 

In our model, we have not taken into account the effect of water on the polaron.
Recent studies~\cite{Kravec2011,Thazhathveetil2011} have suggested that 
its effect would be to reduce the polaron size, which, as can be seen from 
equation (\ref{eq_phic}), corresponds to increasing the value of $g$. 
The net effect would thus be to 
move DNA to a parameter region where the polaron is stronger, as can be seen 
in figures \ref{Fig2} and \ref{Fig4}. As a result the polaron would then 
be able to sustain smaller loops. 

\section{Polyacetylene}
\label{Sec:Poly}
For polyacetylene, the physical value of the parameters are given by
$R_0\approx 0.24nm$,
$W\approx2.5eV$, $\hat{\sigma}\approx21eV/A^2$~\cite{Heeger88},
$\hat{k}\approx 3.7eV/$~\cite{VanNice84} and so
\begin{eqnarray}
\begin{array}{llll}
g \approx 2.56 &\sigma \approx 128 & k\approx 3.79  &M\approx3.2\times 10^4 
\nonumber\\
 k_BT\approx 0.026 & 
\Gamma \approx 1316 &
\tau_0\approx 6.74\times 10^{-16}s&
\end{array}
\end{eqnarray}
Once again, we took $d=1$ and $\alpha=2$ to avoid the potentially 
spurious effects induced at close distance by $J_{n,m}$. 
In this case, we were able to make loops out of $12$ nodes and hairpin-loops
of $12$ nodes too.

Under thermal fluctuation, both were very stable. In this case, 
the integration time 
step was approximately $7\,10^{-18}s$ and our attempt to evaluate the 
configuration average life time was achieved by running 100 simultaneous each
performing over $10^{10}$ integration steps and we did not observe 
a single unfolding of the chain in $13\mu s$. We also ran simulations
for an hairpin-loop configuration of $N=18$ nodes and $L=12$ and also did not 
observe a single unfolding of the chain in over $10\mu s$.

\section{Conclusions}
\label{Sec:Conclusions}
In this paper we have studied the possibility of a polaron to sustain loops 
and hairpin-loop configurations. In these configurations the polaron was 
localised over lattice nodes that were well separated along the chain
backbone but spatially close to each other because of the bending of the chain.
The polaron then acted as a linker between the two regions of the chain and 
so could sustain the loop configuration if the chain was not too rigid.  
The Mingaleev model we have used to describe this property takes into account 
the long distance interactions between the electron and the phonon field, 
with a strength decreasing with the distance. For the configurations
we have studied, the most important contribution comes from lattice nodes
that are spatially close to each other and the energy contribution from
next to nearest neighbour is not essential, unlike in our study of 
spontaneous polaron
displacements~\cite{CPZ12} where the next to nearest neighbour terms 
were essential for the polaron to move along the bending gradient of the chain.

We have determined the critical value of the chain rigidity $k_{crit}$ as 
a function of the polaron coupling constant $g$ and we have shown that 
polarons are able to sustain relatively small loops for a wide range of 
parameters values.

We have then shown that both DNA and polyacetylene are flexible enough
for a polaron to sustain hairpin-loop configurations. Moreover, 
we have also shown that these hairpin-loop configurations are very stable 
under thermal excitations,
with average live times exceeding $10\mu$s, and that they can facilitate the
formation of hairpin-loops of single stranded DNA.

\section{Acknowledgements}
{ BC was partially supported by EPSRC grant EP/I013377/1. 
}

{}

\end{document}